# HERITRACE in action: the ParaText project as a case study for semantic data management in Classical Philology

Francesca Filograsso, Arcangelo Massari, Camillo Neri, Silvio Peroni

## Abstract

HERITRACE is a semantic data editor designed for cultural heritage institutions, addressing the gap between complex Semantic Web technologies and domain expert needs. ParaText Bibliographical Database, a specialized bibliographical database for ancient Greek exegesis, demonstrates HERITRACE's capabilities in Classical Philology. This paper examines how HERITRACE enables non-technical scholars to manage complex semantic data through SHACL-based form generation and validation, while ensuring comprehensive provenance tracking and change management via an OpenCitations Data Model adaptation.

## Introduction

In the digitisation era, classical philologists face a fundamental challenge: managing complex bibliographic data that requires both domain expertise and technical precision. Semantic data - structured information represented using Resource Description Framework (RDF) standards that enables machines to understand relationships between concepts (Manola and Miller 2004) - offers powerful solutions for representing scholarly knowledge, but its technical complexity has traditionally excluded humanities experts from direct data curation.

The ParaText project exemplifies this challenge. ParaText aims to investigate relationships between literary texts and exegetical paratexts in manuscript transmission of ancient Greek poetry, develop hypertextual transcription tools for scholars, and enhance Italian library heritage through public exhibitions of medieval manuscripts. The project produces a *Repertoire of Hypertext Transcriptions*, a *Lexicon of Ancient Greek Exegesis*, and a critical *Bibliographical Database* - all requiring sophisticated semantic data management beyond traditional database approaches[1].

ParaText is a two-year research project (PRIN 2022: October 2023-October 2025) funded through the European Union's Next Generation EU program and the Italian Ministry of University and Research (MUR). The project represents a collaborative effort involving multiple Italian universities: the University of Pavia serves as the coordinating institution, with participating units from the University of Bologna (Department of Classical Philology and Italian Studies), Università Cattolica del Sacro Cuore, University of Genoa, University of Messina, and associated partnerships with the University of Milan, Roma Tor Vergata and eCampus University. This inter-institutional framework was established specifically to address the technical and methodological challenges that individual philological expertise cannot resolve independently.

The complexity of managing semantic data in Classical Philology stems from the discipline's distinctive scholarly requirements. Scholars work with highly specialized terminology, contested interpretations that evolve through scholarly discourse, and complex relationships between texts, commentaries, and manuscript traditions. Traditional database systems cannot adequately represent these nuanced relationships, while existing Semantic Web tools require technical expertise that most humanities scholars lack.

HERITRACE bridges this gap by providing a user-friendly semantic data editor that conceals RDF complexity while preserving semantic integrity (Massari and Peroni 2025). The system enables domain experts to create, edit, and maintain semantically valid datasets with comprehensive provenance tracking, focusing on data curation rather than technical implementation.

---

[1] https://paratext.unipv.it/

# The domain challenge

Classical Philology's semantic data requirements become evident through comparison between specialized and generalist databases. The Année Philologique, founded by Jules Marouzeau in 1926 (Hilbold 2018), serves as the primary bibliographical reference for classical studies with over 900,000 records. However, its generalist approach cannot capture the nuanced distinctions required for specialized research in ancient Greek exegesis.

To cite just one significant example, the subject 'Scholia, commentaries' in Année Philologique encompasses 521 bibliographic records representing a broad category insufficient for detailed scholarly investigation. ParaText employs far more specific terminology: while both databases use 'scholia', ParaText distinguishes it from 'hypomnema' (commentary) and provides granular classifications including 'D-scholia', 'VMK-scholia', 'scholia exegetica', 'h-scholia', and 'Ge-scholia' for different Homeric scholiastic classes. Additional distinctions include temporal categories like 'scholia vetera' for ancient scholia and typological specifications such as 'short scholia', 'frame-scholia', and 'scholia à recueil' indicating length and placement in the manuscript.

Among approximately 400 records pertaining to ancient Greek poetic text exegesis in ParaText Bibliographical Database, 179 employ the 'scholia' keyword. Considering that among Année Philologique's 521 'Scholia, commentaries' records, only few pertain to ancient Greek poetry, many resources remain hidden from potentially interested scholars. This demonstrates that specialized terminological precision enables research depth impossible through generalist approaches.

ParaText Bibliographical Database organizes this complexity through four hierarchical macro-categories: 'exegetical cultures and activities', 'exegetical products', 'exegetical signs and layout', and 'ancient tradition'. Each macro-category contains specialized terms, and when any term is entered as a keyword, the corresponding macro-category must also be included, enabling research at both general and detailed levels while maintaining controlled vocabulary standards.

The scholarly tradition of questioning transmitted texts directly influences bibliographical databases in Classical Philology. This critical approach manifests in specialized record types like 'Review Article' (modelled as `fabio:ReviewArticle` from the FaBiO ontology) alongside standard 'Review' category, and 'in response to' link types highlighting scholarly debates (Peroni and Shotton 2012). However, semantic data face the most scrutiny as they result from critical editorial decisions that directly impact research effectiveness.

# HERITRACE solution

HERITRACE addresses these challenges through a comprehensive semantic data management platform designed specifically for cultural heritage applications. The system employs SHACL (Shapes Constraint Language) - a W3C standard for validating RDF data - to automatically generate user interfaces and validate semantic consistency without requiring users to understand underlying RDF structures (Pareti and Konstantinidis 2022).

SHACL defines constraints and validation rules for semantic data models, specifying which properties entities can have, their data types, and relationships between entities. HERITRACE interprets these SHACL definitions to automatically generate appropriate form interfaces: text fields for literal values, dropdown menus for controlled vocabularies, and relationship selectors for linked entities. This approach ensures that users work within semantically valid boundaries while using familiar form-based interfaces.

For ParaText Bibliographical Database, HERITRACE implements an adaptation of the OpenCitations Data Model (OCDM), a comprehensive ontological framework designed for bibliographic metadata representation in scholarly contexts (Daquino et al. 2020). OCDM extends established Semantic Web vocabularies including FaBiO for publication types (Peroni and Shotton 2012), DataCite for identifier management, and FRBR for hierarchical relationships between bibliographic entities. This adaptation enables sophisticated representation of Classical Philology resources while maintaining compatibility with broader scholarly infrastructure.

The system's provenance model addresses scholarly requirements for comprehensive change tracking and accountability. Each entity modification generates timestamped records capturing essential metadata: temporal information indicating when changes occurred, agent information linking modifications to authenticated users through ORCID integration, and source documentation connecting changes to evidential basis. This provenance tracking operates transparently, requiring no additional effort from users while ensuring complete audit trails.

HERITRACE implements change management through snapshot-based versioning, storing complete entity states at each modification point (Peroni et al. 2016). This approach enables scholars to examine historical versions, compare changes over time, and restore previous states when necessary.

The system integrates authentication through ORCID (Open Researcher and Contributor ID), ensuring that all changes are attributed to verified scholarly identities. This integration supports institutional accountability requirements while enabling collaboration across organizations.

## Results and impact

ParaText Bibliographical Database implementation using HERITRACE demonstrates advantages over existing approaches to bibliographic data management in specialized humanities contexts. The system enables non-technical scholars to create and maintain semantic datasets while preserving the interpretive flexibility essential for humanities research.

The case study of Chiara Martis's 2013 article *L'enigma del PLouvre inv. 7733 verso: l'epigramma dell'ostrica* illustrates how semantic precision impacts research effectiveness (Martis 2013). Comparison of abstracts across three sources reveals editorial challenges in terminology selection. The journal refers to "explanatory commentary" and acknowledges an "elegy or epigram" debate (fig. 1), Année Philologique refers generically to "distici elegiaci" and "commento" (fig. 2), while ParaText Bibliographical Database initially used "commented edition" – the most precise term identifying the exegetical product as text-plus-paratext unity rather than paratext alone – but referred only to "epigram" (fig. 3).

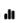

**Figure 1**: Abstract from *Studi di Egittologia e di Papirologia* 10 (2013)

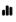

**Figure 2**: Bibliographic record from Année Pihlologique database

> **Title**
>
> L'enigma del PLouvre inv. 7733 verso: l'epigramma dell'ostrica
>
> **Abstract**
>
> P. Louvre 7733 is a commented edition which contains paragraphoi enhanced by vacua to mark pause or transition to another topic, expunction marks and interlinear emendations; the text of the poem, an epigram preferably dated between the 1st century BC and the 1st century AD, is examined together with the commentary of which are available the text and the Italian translation

Figure 3: Abstract from ParaText database

However, ParaText Bibliographical Database's initial choice to use only 'epigram' in the abstract and as keyword concealed ongoing scholarly debate and limited discoverability for researchers studying paratexts related to elegy. This necessitated scholarly discussion, ultimately leading to abstract revision and introduction of the 'elegy' keyword to maintain research accessibility while preserving interpretive debate – demonstrating how well-designed semantic systems can enhance rather than constrain scholarly discourse (fig. 4).

**Title**

L'enigma del PLouvre inv. 7733 verso: l'epigramma dell'ostrica

**Abstract**

P. Louvre 7733 is a commented edition which contains paragraphoi enhanced by vacua to mark pause or transition to another topic, expunction marks and interlinear emendations; the text of the poem, an epigram rather than an elegy, is preferably dated between the 1st century BC and the 1st century AD and examined together with the commentary of which are available the text and the Italian translation

Figure 4: Revised abstract from *ParaText* database

The system's accommodation of contested interpretations proves particularly valuable for humanities applications where multiple valid perspectives may coexist and evolve through scholarly discourse. ParaText Bibliographical Database editors can modify semantic classifications and abstracts as research progresses, while maintaining complete historical records of changes and their justifications.

# Conclusion and future directions

HERITRACE demonstrates that semantic data management can support specialized humanities scholarship while preserving scholarly rigor and accessibility for non-technical users. The ParaText Bibliographical Database implementation provides evidence that carefully designed semantic systems can accommodate sophisticated Classical Philology requirements including contested interpretations, evolving terminology, and complex bibliographic relationships.

However, the ParaText Bibliographical Database project reveals significant limitations arising from interdisciplinary collaboration without preliminary data model coordination. The project emerged from collaboration between distinct research teams - expert in text and transmission of ancient Greek poetry and related exegesis from Antiquity to the Byzantine period from the Department of Classical Philology and Italian Studies at the University of Bologna, and computer scientists from OpenCitations - each pursuing independent objectives. This distributed approach, while valuable for maintaining disciplinary autonomy, precluded essential preliminary phases of data model definition tailored to Classical Philology requirements.

Consequently, ParaText Bibliographical Database adapted the existing OpenCitations Data Model, originally designed for bibliographic metadata of scientific publications, rather than developing a purpose-built ontology for classical studies. While OCDM provides robust foundations for scholarly metadata, its scientific publication focus cannot fully capture the sophisticated semantic requirements of Classical Philology, particularly specialized terminological hierarchies that characterize ancient Greek scholarship.

Additionally, future developments should address the integration of HERITRACE with presentation-oriented platforms such as OSCAR (Heibi et al. 2018), specifically designed for data visualization and presentation. While HERITRACE focuses on semantic data curation and management, the complete scholarly workflow requires complementary tools for effective data dissemination and user engagement with the curated semantic dataset.

# Acknowledgment

This research was supported by the Italian Ministry of University and Research (MUR) through the PRIN 2022 program "Projects of Relevant National Interest" (Progetti di Ricerca di Rilevante Interesse Nazionale), project code 2022NE2LWP, co-funded by the European Union under the Next Generation EU initiative within the framework of the National Recovery and Resilience Plan (PNRR) - Mission 4 "Education and Research", Component 2 "From Research to Business", Investment 1.1.

Special thanks go to each member of the ParaText team for their suggestions, which have contributed significantly to enhance the bibliographical database.

# Author contributions

Using the CRediT[2] (Contributor Roles Taxonomy):
- Francesca Filograsso: Data curation, Writing - Original Draft
- Arcangelo Massari: Software, Methodology, Writing - Original Draft
- Camillo Neri: Supervision, Funding Acquisition, Writing - Review & Editing
- Silvio Peroni: Supervision, Funding Acquisition, Writing - Review & Editing

# References

Daquino, Marilena, Silvio Peroni, David Shotton, et al. 2020. "The OpenCitations Data Model." In *The Semantic Web – ISWC 2020*, edited by Jeff Z. Pan, Valentina Tamma, Claudia d'Amato, et al., vol. 12507. Lecture Notes in Computer Science. Springer International Publishing. https://doi.org/10.1007/978-3-030-62466-8_28.

Heibi, Ivan, Silvio Peroni, and David Shotton. 2018. "OSCAR: A Customisable Tool for Free-Text Search over SPARQL Endpoints." In *Semantics, Analytics, Visualization*, edited by Alejandra González-Beltrán, Francesco Osborne, Silvio Peroni, and Sahar Vahdati, vol. 10959. Lecture Notes in Computer Science. Springer International Publishing. https://doi.org/10.1007/978-3-030-01379-0_9.

Hilbold, Ilse. 2018. "Jules Marouzeau et « L'Année Philologique »: Aux Origines de La Réforme de La Bibliographie d'études Classiques." *Revue Des Études Latines* 96: 239–58.

Manola, F., and E. Miller. 2004. "RDF Primer." W3C, February 10. http://www.w3.org/TR/2004/REC-rdf-primer-20040210/.

Martis, Chiara. 2013. "L'enigma del PLouvre inv. 7733 verso : l'epigramma dell'ostrica." *Studi di egittologia e papirologia : 10, 2013*, 117–50. https://doi.org/10.1400/213891.

Massari, Arcangelo, and Silvio Peroni. 2025. "HERITRACE: A User-Friendly Semantic Data Editor with Change Tracking and Provenance Management for Cultural Heritage Institutions." *Umanistica Digitale*, no. 20 (July): 20. https://doi.org/10.6092/issn.2532-8816/21218.

Pareti, Paolo, and George Konstantinidis. 2022. "A Review of SHACL: From Data Validation to Schema Reasoning for RDF Graphs." In *Reasoning Web.*

---

[2] https://credit.niso.org/

*Declarative Artificial Intelligence*, edited by Mantas Šimkus and Ivan Varzinczak, vol. 13100. Lecture Notes in Computer Science. Springer International Publishing. https://doi.org/10.1007/978-3-030-95481-9_6.

Peroni, S., D. Shotton, and F. Vitali. 2016. "A Document-Inspired Way for Tracking Changes of RDF Data." In *Detection, Representation and Management of Concept Drift in Linked Open Data*, edited by L. Hollink, S. Darányi, A.M. Peñuela, and E. Kontopoulos. CEUR Workshop Proceedings. http://ceur-ws.org/Vol-1799/Drift-a-LOD2016_paper_4.pdf.

Peroni, Silvio, and David Shotton. 2012. "FaBiO and CiTO: Ontologies for Describing Bibliographic Resources and Citations." *Journal of Web Semantics* 17 (December): 33–43. https://doi.org/10.1016/j.websem.2012.08.001.